\documentclass{LMCS}

\usepackage{amssymb,amsmath, stmaryrd}
\usepackage{program}
\usepackage{graphicx,psfig,epsf}
\usepackage[all]{xy}
\usepackage{hyperref}

\newcommand{\myarrow}[1]{\stackrel{#1}{\longrightarrow}}
\newcommand{\notarrow}[1]{\not\stackrel{\, \, #1}{\longrightarrow}}
\newcommand{\must}[1]{\stackrel{\mathit{#1}}{\longrightarrow}_{\mathit{must}}}
\newcommand{\mustminus}[1]{\stackrel{\mathit{#1}}{\longrightarrow}_{\mathit{must}^-}}
\newcommand{\mustplus}[1]{\stackrel{\mathit{#1}}{\longrightarrow}_{\mathit{must}^+}}
\newcommand{\may}[1]{\stackrel{\mathit{#1}}{\longrightarrow}_{\mathit{may}}}
\newcommand{\RL}{RL}
\newcommand{\algname}{\alpha\textsc{Search}}
\newcommand{\refalgname}{\textsc{RefinementSearch}}
\newcommand{\infm}{\mathit{Inf}_M}
\newtheorem{lemma}{Lemma}
\newtheorem{theorem}{Theorem}
\newtheorem{corollary}{Corollary}
\newtheorem{definition}{Definition}
\newcommand{\semantics}[1]{\llbracket #1 \rrbracket}
\newcommand{\mustaa}{\mathit{must}}

\def\doi{3 (1:5) 2007}
\lmcsheading%
{\doi}
{1--22}
{}
{}
{Feb.~\phantom{0}3, 2006}
{Feb.~26, 2007}
{}

\begin{document}

\title[Under-Approximation Refinement]{Predicate Abstraction With\\
Under-Approximation Refinement\rsuper *}

\author[C.~S.~P\u{a}s\u{a}reanu]{Corina S. P\u{a}s\u{a}reanu\rsuper a}
\address{{\lsuper a}QSS, NASA Ames Research Center, Moffett Field, CA 94035, USA}
\email{pcorina@email.arc.nasa.gov}

\author[R.~Pel\'anek]{Radek Pel\'anek\rsuper b}

\address{{\lsuper b}Masaryk University Brno, Czech Republic}
\email{xpelanek@fi.muni.cz}
\thanks{{\lsuper b}Partially supported by the Grant Agency of Czech
Republic grant No. 201/07/P035 and by the Academy of Sciences of Czech
Republic grant No. 1ET408050503.}

\author[W.~Visser]{Willem Visser\rsuper c}
\address{{\lsuper c}RIACS/USRA, NASA Ames Resarch Center, Moffett Field, CA 94035, USA}
\email{wvisser@email.arc.nasa.gov}

\titlecomment{{\lsuper *}This paper is an extended version of~\cite{PasareanuPelanekVisserCAV05}.}


\begin{abstract}
  \noindent We propose an abstraction-based model checking method which 
  relies on refinement of an under-approximation of the feasible behaviors 
  of the system under analysis. 
  The method preserves errors to safety properties, since all
  analyzed behaviors are feasible by definition.  The method does not require
  an abstract transition relation to be generated, but instead executes the
  concrete transitions while storing abstract versions of the concrete states,
  as specified by a set of abstraction predicates.
  For each explored transition the method checks, with the help of a theorem 
  prover, whether there is any loss of precision introduced by abstraction.
  The results of these checks are used to decide termination or to refine the 
  abstraction by generating new abstraction predicates.
  If the (possibly infinite) concrete system under analysis has a finite
  bisimulation quotient, then the method is guaranteed to eventually explore an
  equivalent finite bisimilar structure. We illustrate the application of the 
  approach for checking concurrent programs. 
\end{abstract}

\maketitle

\section{Introduction}

Over the last few years, model checking based on
abstraction-refinement has become a popular technique for the
analysis of software. In particular the abstraction technique of
choice is a property preserving over-approximation called
predicate abstraction~\cite{GrafSaidi97} and the refinement
removes spurious behavior based on automatically analyzing
abstract counter-examples. This approach is often referred to as
CEGAR (counter-example guided automated refinement) and forms the
basis of some of the most popular software model checkers
\cite{759192,MAGIC,BLAST-IDEA}.
Furthermore, a strength of model checking is its ability to automate the
detection of subtle
errors and to produce traces that exhibit those errors. However,
over-approximation based abstraction techniques are not particularly well
suited for this, since the detected defects may be spurious due to the
over-approximation --- hence the need for refinement. We propose an alternative
approach based on refinement of under-approximations,
which effectively preserves the defect detection ability of model
checking in the presence of aggressive abstractions.

The technique uses a combination of (explicit state) model
checking, predicate abstraction and automated refinement to
efficiently analyze increasing portions of the feasible behavior
of a system. At each step, either an error is found, we are
guaranteed no error exists, or the abstraction is refined. More
precisely, the proposed model checking technique traverses the
concrete transitions of the system and for each explored concrete
state, it stores an abstract version of the state. The abstract
state, computed by predicate abstraction, is used to determine
whether the model checker's search should continue or backtrack
(if the abstract state has been visited before). This effectively
explores an under-approximation of the feasible behavior of the
analyzed system. Hence all counter-examples to safety properties
are preserved.

Refinement uses weakest precondition calculations to check, with
the help of a theorem prover, whether the abstraction introduces
any loss of precision with respect to each explored transition.
If there is no loss of precision due to
abstraction (we say that the abstraction is {\em exact}) the search stops and
we conclude that the property holds.
Otherwise, the results from the failed checks are used to refine the
abstraction and the whole verification process is repeated anew.
In general, the iterative refinement may not 
terminate. However, if a finite bisimulation quotient~\cite{LEE92}
exists for the system under analysis, then the proposed approach
is guaranteed to eventually explore a finite structure that is
bisimilar to the original system.

The technique can also be
used in a lightweight manner, without a theorem prover, i.e. the
refinement guided by the exactness checks is replaced with
refinement based on syntactic substitutions \cite{734102}
or heuristic refinement.
The proposed technique can be used for systematic testing, as it
examines increasing portions of the system  under analysis. In
fact, our method extends existing approaches to
testing that use abstraction mappings~\cite{ASML,ROSTRA}, by adding
support for automated abstraction refinement.

Our approach can be contrasted with the work on predicate abstraction for modal
transition systems~\cite{701647,ShohamG04}, used in the
verification and refutation of branching time temporal logic properties. An
abstract model for such logics distinguishes between {\em may} transitions,
which over-approximate transitions of the concrete model, and {\em must}
transitions, which under-approximate the concrete transitions (see
also~\cite{ball04,DamsN04,AlfaroGJ04,ShohamG03}).
As we show in the next section (and we discuss in more detail in 
Section~\ref{sec:discussions}), the technique presented here explores 
and generates a structure which is more  precise (contains more feasible 
behaviors) than the model defined by the
{\em must} transitions, for the same abstraction predicates.  The reason is
that the model checker explores transitions that correspond not only to {\em
must} transitions, but also to {\em may} transitions that are feasible. 

Moreover, unlike~\cite{701647,ShohamG04} and
over-approximation based abstraction techniques \cite{759192,MAGIC}, the
under-approximation and refinement approach does not require the a priori
construction of the abstract transition relation, which involves exponentially
many theorem prover calls (in the number of predicates), regardless of the size
of (the reachable portion of) the analyzed system. In our case, the model
checker executes concrete transitions and a theorem prover is only used during
refinement, to determine whether the abstraction is exact with respect to each
executed transition. Every such calculation makes at most two theorem prover
calls, and it involves only the {\em reachable} state space of the system under
analysis. Another difference with previous abstraction techniques is that the
refinement process is not guided by the spurious counter-examples, since no
spurious behavior is explored. Instead, the refinement is guided by the failed
exactness checks for the explored transitions.

To the best of our knowledge, the presented approach is the first
predicate abstraction based analysis which focuses on automated refinement of
under-approximations with the goal of efficient error detection.
We illustrate the application of the
approach for checking safety properties in concurrent programs.

The rest of the paper is organized as follows.
Section~\ref{sec:example} shows an example illustrating our
approach. Section~\ref{sec:background} gives background
information. Section~\ref{sec:algorithm} describes the main
algorithm for performing concrete model checking with abstract
matching and refinement. Section~\ref{sec:correctnesstermination} discusses
correctness and termination; Section~\ref{sec:discussions} discusses other interesting properties for 
the algorithm.
Section~\ref{sec:extensions} proposes extensions to the 
algorithm. Section~\ref{sec:applications} illustrates applications
of the approach, Section~\ref{sec:related-work} discusses related work, 
and  Section~\ref{sec:conclusions} concludes the paper.

\section{Example}
\label{sec:example}
\begin{figure}[!t]
\epsfxsize=4.9in
\centerline{\epsffile{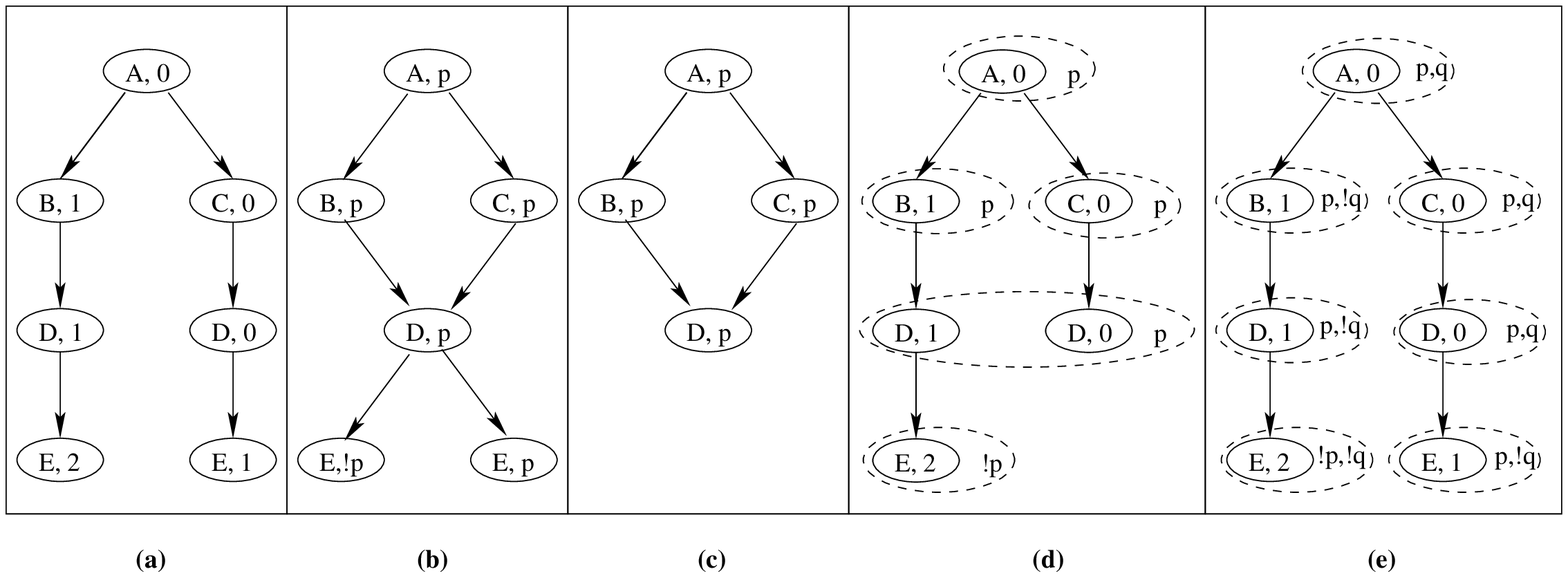}}
\caption{{\bf (a)} Concrete system  {\bf (b)} {\em May} abstraction using predicate $p=x<2$
{\bf (c)} {\em Must} abstraction using $p$
{\bf (d)} Concrete search with abstract matching using $p$
{\bf (e)} Concrete search with abstract matching using predicates $p$ and $q=x<1$}
\label{fig:comparison}

\end{figure}
The example in Figure~\ref{fig:comparison} illustrates some of the
main characteristics of our approach. Figure~\ref{fig:comparison}
{\bf (a)} shows the state space of a concrete system that has only
one variable $x$; states are labelled with the program counter
(e.g. $A$, $B$, $C$, \ldots) and the concrete value of~$x$.
Figure~\ref{fig:comparison} {\bf (b)} shows the abstract system induced by the
{\em may} transitions for predicate $p=x<2$.
Figure~\ref{fig:comparison} {\bf (c)} shows the abstract system induced by the
{\em must} transitions for predicate $p$.

Figure~\ref{fig:comparison} {\bf (d)} shows the state space
explored using our proposed approach, for an abstraction specified
by predicate $p$. Dotted circles denote the abstract states which
are stored, and used for matching, during the concrete execution
of the system. The approach explores only the {\em feasible}
behavior of the concrete system, following transitions that
correspond to both {\em may} and {\em must} transitions, but it
might miss behavior due to abstract matching. For example, 
state 
$(E,1)$ is not explored, assuming a
breadth-first search, since $(D,0)$ was matched with $(D,1)$ ---
both have the same program counter and both satisfy $p$. 
Notice that, with respect to reachable states, the produced structure is a
better under-approximation (it ``covers'' more states) than the {\em must} 
abstraction.
Figure~\ref{fig:comparison} {\bf (e)} illustrates concrete
execution with abstract matching, after a refinement step, which
introduced a new predicate $q=x<1$. The resulting structure is an
exact abstraction of the concrete system.

\section{Background}
\label{sec:background}

\subsection{Program Model and Semantics}
To make the presentation simple, we use as a specification language a guarded
commands language over integer variables. Most of the results extend directly
to more sophisticated programming languages. Let $V$ be a finite set of integer
variables. Expressions over $V$ are defined using standard boolean ($=, <,>$)
and binary ($+, -, \cdot, ...$) operations.

\begin{definition}
  A \emph{model} is a tuple $M=(V, T)$. $T = \{ t_1, \ldots, t_k \}$ is a
  finite set of transitions, where $t_i = (g_i(\vec{x})\longmapsto \vec{x} :=
  e_i(\vec{x}))$, $g_i(\vec{x})$ is a guard and $e_i(\vec{x})$ are assignments
  to the variables represented by tuple $\vec{x}$.
\end{definition}

Throughout the paper, we write concurrent assignments 
$\vec{x} := e_i(\vec{x}))$ as sequences, 
to improve readability. The semantics of program models uses transition 
systems.

\begin{definition}
  A \emph{transition system} over a finite set of atomic propositions $AP$ is a
  tuple $(S, R, s_0, L)$ where $S$ is a (possibly infinite) set of states, $R =
  \{ \myarrow{i} \}$ is a finite set of deterministic transition relations:
  $\myarrow{i} \subseteq S\times S$, $s_0$ is an initial state, and $L: S
  \rightarrow 2^{AP}$ is a labeling function.
\end{definition}

State $s$ is {\em reachable} if there exists a sequence of zero or more
transitions from the initial state such that $s_0 \myarrow{i_1} s_1
\myarrow{i_2} s_2 ... \myarrow{i_n} s_n = s$ (denoted $s_0 \myarrow{}^* s$).
The set of \emph{reachable labelings} $ \RL(T)$ is $ \{ L(s) \mid s\in S: s_0
\myarrow{}^* s \}$.
The notation $s \notarrow{i}$ means that there is no $i$ transition from the
state $s$. 

\begin{definition}
\label{def:concretesemantics}
The \emph{concrete semantics} of model $M$ is transition system $\semantics{M}$
  $=$ $(S$, $\{\myarrow{i}\}$, $s_0,L)$ over $AP$, where:
\begin{itemize}
\item $S = 2^{V\rightarrow \mathbb{Z}}$, i.e. states are valuations of
  variables,
\item $s \myarrow{i} s' \Leftrightarrow s\models g_i \wedge s' = e_i(s)$;
  the semantics of guards (boolean expressions) and updates is as usual;
  guards are functions $(V\rightarrow \mathbb{Z}) \rightarrow \{ \mathit{true},
  \mathit{false} \}$, written as $s \models g_i$; updates are functions
  $e_i:(V\rightarrow\mathbb{Z})\rightarrow(V\rightarrow\mathbb{Z})$,
\item $s_0$ is the zero valuation ($\forall v\in V: s_0(v) = 0$),
\item $L(s) = \{ p \in AP \mid s \models p \}$.
\end{itemize}
\end{definition}

\subsection{Strongest Postcondition and Weakest Precondition}
Let $\phi$ be a predicate representing a set of states. Then the {\em strongest postcondition}
of $\phi$ with respect to transition $i$ is $sp(\phi,i)=\exists s'.(s'\myarrow{i} s \wedge\phi(s'))$;
$sp(\phi,i)$ defines the successors by transition $i$ of the states characterized by $\phi$.
The {\em weakest precondition} of $\phi$ with respect to transition $i$ is 
$wp(\phi,i)=\forall s'.(s\myarrow{i} s' \Rightarrow \phi(s'))$; $wp(\phi,i)$ characterizes 
the largest set of states whose successors by transition $i$ satisfy $\phi$. For guarded commands,
the weakest precondition can be expressed as $wp(\phi, i) = (g_i\Rightarrow
\phi[e_i(\vec{x})/\vec{x}])$. We will use the following property~\cite{GrafSaidi97}: $sp(\phi,i)\Rightarrow \phi'$ iff
$\phi \Rightarrow wp(i,\phi')$.

\subsection{Predicate Abstraction}
Predicate abstraction is a special instance of the framework of abstract
interpretation \cite{COU92} that maps a (potentially infinite state) transition
system into a finite state transition system via a set of predicates $\Phi =
\{\phi_1, \ldots, \phi_n \}$ over the program variables. Let $\mathbb{B}_n$ be
a set of bitvectors of length~$n$. We define abstraction function $\alpha_\Phi:
S \rightarrow \mathbb{B}_n$, such that $\alpha_\Phi(s)$ is a bitvector
$b_1b_2\ldots b_n$ such that $b_i =1 \Leftrightarrow s \models \phi_i$. Let
$\Phi_s$ be the set of all abstraction predicates that evaluate to {\em true}
for a given state $s$, i.e. $\Phi_s=\{\phi\in\Phi \mid s \models \phi\}$. For
succinctness we sometimes write $\alpha_\Phi(s)$ to
denote $\bigwedge_{\phi\in\Phi_s}
\phi\wedge\bigwedge_{\phi\notin\Phi_s}\neg\phi$.

We also give here the definitions of {\em may} and {\em must} abstract
transitions. Although not necessary for formalizing our algorithm, these
definitions clarify the comparison with related work. For two abstract states
(bitvectors) $a_1$ and $a_2$:
\begin{itemize}
\item $a_1 \must{i} a_2$ iff for all concrete states $s_1$ such that $\alpha_\Phi(s_1)=a_1$,
there exists concrete state $s_2$ such that $\alpha_\Phi(s_2)=a_2$ and $s_1\myarrow{i} s_2$,
\item $a_1 \may{i} a_2$ iff there exists concrete state $s_1$ such
  that $\alpha_\Phi(s_1)=a_1$ and there exists concrete state $s_2$ such that
  $\alpha_\Phi(s_2)=a_2$, such that $s_1\myarrow{i} s_2$.
\end{itemize}

Algorithms for computing abstractions using over-approximation
based predicate abstraction are given in e.g.
\cite{759192,GrafSaidi97} (they compute {\em may} abstract
transitions automatically, with the help of a theorem prover). In
the worst case, these algorithms make $2^n\times n\times 2$ calls
to the theorem prover for each program transition. 

\subsection{Bisimulation}

\begin{definition}  
  A symmetric relation $R\subseteq S\times S$ is a \emph{bisimulation relation}
  iff for all $(s, s') \in R$:
\begin{itemize}
\item $L(s) = L(s')$
\item For every $s \myarrow{i} s_1$ there exists $s' \myarrow{i} s_1'$ such
  that $R(s_1,s_1')$
\end{itemize}
\end{definition}

The \emph{bisimulation} is the largest bisimulation relation, denoted $\sim$.
Two transition systems are bisimilar if their initial states are bisimilar. As
$\sim$ is an equivalence relation, it induces a {\em quotient} transition
system whose states are equivalence classes with respect to $\sim$ and there is
a transition between two equivalence classes $A$ and $B$ if $\exists s_1\in A$
and $\exists s_2\in B$ such that $s_1 \myarrow{i} s_2$.

\section{Concrete Model Checking with Abstract Matching}
\label{sec:algorithm}

\subsection{Algorithm}
Figure~\ref{fig:search-with-check} shows the reachability procedure
that performs model checking with abstract matching ($\algname$).
It is basically concrete state space exploration with matching on
abstract states; the main modification with respect to classical
state space search is that we store $\alpha_\Phi(s)$ instead of $s$.
The procedure uses the following data structures:
\begin{itemize}
\item $|States|$ is  a set of abstract states visited so far,
\item $|Transitions|$ is a set of abstract transitions visited so far,
\item $|Wait|$ is a set of concrete states to be explored.
\end{itemize}

 \begin{figure}[t]
   \begin{center}
\begin{program}
\PROC \algname(M, \Phi)
 \Phi_{new} = \Phi; \mbox{add } s_0 \mbox{ to } |Wait|; \mbox{add } \alpha_\Phi(s_0) \mbox{ to } |States|
 \WHILE |Wait| \neq \emptyset \DO
        \mbox{get } s \mbox{ from } |Wait|
        L(\alpha_\Phi(s)) = \{ a \in AP \mid s \models a \}
        \FOREACH i \mbox{ from } 1 \mbox{ to } n \DO
                \IF s \models g_i \THEN
                        \IF \alpha_\Phi(s) \Rightarrow g_i \mbox{ is not valid}
                         \THEN \mbox{add } g_i \mbox{ to } \Phi_{new} \FI
                        s' = e_i(s)
                        \IF \alpha_\Phi(s) \Rightarrow \alpha_\Phi(s')[e_i(\vec{x})/\vec{x}] \mbox{ is not valid}
                         \THEN \mbox{add predicates in } \alpha_\Phi(s')[e_i(\vec{x})/\vec{x}] \mbox{ to }  \Phi_{new} \FI
                        \IF \alpha_\Phi(s') \not \in |States| \THEN
                        \mbox{add } s' \mbox{ to } |Wait|
                        \mbox{add } \alpha_\Phi(s') \mbox{ to } |States|
                        \FI
                        \mbox{add } (\alpha_\Phi(s), i, \alpha_\Phi(s')) \mbox{ to } |Transitions|

                \ELSE
                        \IF \alpha_\Phi(s) \Rightarrow \neg g_i \mbox{ is not valid}
                         \THEN \mbox{add } g_i \mbox{ to }  \Phi_{new} \FI
                \FI
        \OD
 \OD
  A = (|States|, |Transitions|, \alpha_\Phi(s_0), L)
 \mbox{return } (A, \Phi_{new})
\END
\end{program}

\caption{Search procedure with checking for exact abstraction}
     \label{fig:search-with-check}

   \end{center}
\end{figure}
The procedure performs validity checking, using a theorem prover, to determine
whether the abstraction is {\em exact} with respect to each explored transition
--- see discussion below. The set $\Phi_{new}$ maintains the list of abstraction
predicates. The procedure returns the computed structure and a set of new
predicates that are used for refinement. Note that we
never abstract the program counter.

Figure~\ref{fig:refinement-search} gives the iterative refinement
algorithm for checking whether $M$ can reach an error state
described by $\varphi$ (which is a boolean combination of propositions from $AP$).  
The algorithm starts with $AP$ as the initial set of abstraction predicates.
At each iteration of the loop, the
algorithm invokes procedure $\algname$ to analyze an
under-approximation of the system, which either violates the
property, it is proved to be correct (if the abstraction is found
to be exact with respect to all transitions), or it needs to be
refined.
Counter-examples are generated as usual (with depth-first search order 
using the stack, with breadth-first search order using parent pointers). 

\begin{figure}[t]
  \begin{center}
\begin{program}
  \PROC \refalgname(M, \varphi)
  j=1; \Phi_j = AP
  \WHILE true \DO
    (A_i, \Phi_{j+1}) = \algname(M, \Phi_j)
    \IF \varphi \mbox{ is reachable in } A_j \THEN \mbox{return counter-example}  \FI
    \IF \Phi_{j+1} = \Phi_j \THEN \mbox{return unreachable} \FI
    j = j+1
  \OD
  \END
\end{program}
    \caption{Iterative refinement algorithm}
    \label{fig:refinement-search}
  \end{center}
\end{figure}

\subsection{Checking for Exact Abstraction and Refinement}
\label{sec:check-exact-abstr}

We say that abstraction function $\alpha_\Phi$ is {\em exact} with respect to
transition $s \myarrow{i} s'$ iff for all $s_1$ such that $\alpha_\Phi(s) =
\alpha_\Phi(s_1)$ there exists $s_1'$ such that $\alpha_\Phi(s_1') =
\alpha_\Phi(s')$ and $s_1 \myarrow{i} s_1'$. In other words, $s \myarrow{i} s'$
is exact with respect to $\alpha_\Phi$ iff $\alpha_\Phi(s) \must{i}
\alpha_\Phi(s')$. 

Moreover, the abstraction function $\alpha_\Phi$ is {\em exact} with respect
to a state $s$ iff the following conditions hold:
(1) $\alpha_\Phi$ is exact with respect to all transitions 
$s \myarrow{i} s'$ and
(2) if $s \notarrow{i}$ then for all $s_1$ such that $\alpha_\Phi(s) =
  \alpha_\Phi(s_1)$ we have $s_1 \notarrow{i}$.

The notion of exactness is related to {\em completeness} in
abstract interpretation (see~\cite{GiacobazziQuintarelli01}), which states
that no loss of precision is introduced by the abstraction.
Checking that the abstraction is exact with respect to a concrete transition
$s\myarrow{i} s'$ amounts to checking that 
$sp(\alpha_\Phi(s)) \Rightarrow \alpha_\Phi(s')$, equivalent to $\alpha_\Phi(s) \Rightarrow wp(\alpha_\Phi(s'),i)$, 
is valid. 

Note that $wp(\alpha_\Phi(s'), i) = 
(g_i\Rightarrow\alpha_\Phi(s')[e_i(\vec{x})/\vec{x}])$.
Therefore 
$\alpha_\Phi(s) \Rightarrow wp(\alpha_\Phi(s'),i)$ is equivalent to
$\alpha_\Phi(s) \Rightarrow (g_i\Rightarrow\alpha_\Phi(s')[e_i(\vec{x})/\vec{x}])$.
The abstraction is exact with respect to state $s$ when the following 
conditions hold:
(1) $\alpha_\Phi(s) \Rightarrow (g_i\wedge
  \alpha_\Phi(s')[e_i(\vec{x})/\vec{x}])$, equivalent to $(\alpha_\Phi(s) \Rightarrow g_i) \wedge
  (\alpha_\Phi(s) \Rightarrow\alpha_\Phi(s')[e_i(\vec{x})/\vec{x}])$, is valid for each $i$ such that $s
  \models g_i$ and
(2)~$\alpha_\Phi(s) \Rightarrow \neg g_i$ is valid for each $i$ such that $s
  \not \models g_i$.

Checking the validity for these formulas is in general undecidable. As is
customary, if the theorem prover can not decide the validity of a formula, we
assume that it is not valid. This may cause some unnecessary refinement, but it
keeps the correctness of the approach. If the abstraction can not be proved to
be exact with respect to some transition, then the new predicates from the
failed formula are added to the set of abstraction predicates. Intuitively,
these predicates will be useful for proving exactness in the next iteration.

\section{Correctness and Termination}
\label{sec:correctnesstermination}

In this section we discuss the main properties of the iterative refinement
algorithm. We first state the main theorems, after which we give the technical lemmas and
proofs (the reader may wish to skip this technical material on the first reading).

\subsection{Main Results}

We first show that, if the iterative algorithm terminates then the result is correct
and moreover, if the error state is unreachable, the output structure is
bisimilar to the system under analysis:

\begin{theorem} (Correctness)
  \label{theorem:correctness}
  If\  $\refalgname(M,\varphi)$ terminates then:
  \begin{itemize}
  \item if it returns a counter-example, then it is a real error, 
  \item if it returns `unreachable', then the error state is indeed unreachable
    in $M$ and moreover the computed structure is bisimilar to $\semantics{M}$.
  \end{itemize}
\end{theorem}

In general, the proposed algorithm might not terminate (the reachability
problem for our modeling language is undecidable). However, the algorithm is
guaranteed to eventually find all the reachable labelings (including all the reachable errors)
of the concrete program, although it might not be able to detect that (to decide termination).
Moreover, if the (reachable part of the) system under analysis has a finite
bisimulation quotient, then the algorithm eventually produces a finite
bisimilar structure.

\begin{theorem} (Termination)
\label{th:termination}
Let the $\algname$ use breadth-first search order and let $A_1$, $A_2$, \ldots be a
sequence of transition systems generated during iterative refinement performed
by $\refalgname(M,\varphi)$. Then
  \begin{itemize}
  \item there exists $j$ such that $RL(A_j) = RL(\semantics{M})$,
\item if the reachable part of the bisimulation quotient is finite, then there
  exists $j$ such that $A_j \sim \semantics{M}$. 
  \end{itemize}
\end{theorem}

Note that a consequence of this theorem is that if an error is reachable it is eventually reported 
by our algorithm. Also note that for the second part of the theorem,
we do not require that both the reachable and unreachable parts of the 
system have a finite bisimulation quotient, but only the reachable part needs to be 
finite (of course, if both the reachable and unreachable parts are finite, then it 
follows that the reachable part is also finite; the converse is not true).

\subsection{Technical Material}

Here we provide several technical lemmas and the proofs for the two main
theorems. We use the following notation: a state $s$ is \emph{visited} during
the search if it is inserted into $|Wait|$; a state $s$ is \emph{considered}
during the search if it is generated as a successor of some state in the
$\FOREACH$ loop; a state $s_1$ is \emph{matched} to a state $s_2$ if the check
$\alpha_\Phi(s_1) \not \in |States|$ fails because $\alpha_\Phi(s_1) =
\alpha_\Phi(s_2)$ and $s_2$ was visited  before.  

We say that transition $s \myarrow{i} s'$ is exact if $\alpha_\Phi$ is exact
with respect to it. Note that sometimes we let $\algname(M,\Phi)$ denote just
the structure $A$ computed by the algorithm and not the tuple $(A,
\Phi_{new})$. Also note that $\refalgname$ starts with $AP$ as the initial set
of predicates. For the proofs, we need to refine the definition of
bisimulation.

\begin{definition}
A symmetric relation $R \subset S\times S$ is a $k$-bisimulation relation iff:
\begin{itemize}
\item for all $(s,s') \in R: L(s) = L(s')$
\item if $k>0$ then there exists $(k-1)$-bisimulation relation $R'$ such that for
  all $(s,s') \in R: (\forall s\myarrow{i} s_1 \Rightarrow \exists s'
  \myarrow{i} s_1' \wedge (s_1,s_1') \in R')$
\end{itemize}
\end{definition}

The $k$-bisimulation is the largest $k$-bisimulation relation, denoted
$\sim_k$. Note that the bisimulation is a $k$-bisimulation relation for every $k$.

\subsubsection*{Proof of Theorem~1}

We first show that the reachable labelings computed by the
iterative algorithm $\refalgname$ is indeed an {\em under-approximation} of the reachable
labelings of the program under analysis (Lemmas~\ref{lemma:must-sequences} and \ref{lemma:must-search-concrete}).
Therefore, all the reported counter-examples correspond to real errors. We then show that when 
 $\refalgname$ reports 'unreachable' (i.e. when the set $\Phi_{new}$ of predicates returned for 
the current iteration is equal to the set $\Phi$ of predicates from the previous iteration)
then the computed structure $A$ is bisimilar to $\semantics{M}$ 
(Lemmas~\ref{lemma:bisim-ok} and \ref{lemma:check-correctness}).

\begin{lemma}
  \label{lemma:must-sequences}
  If a state $s$ is reachable in $\semantics{M}$ via exact transitions with respect to
  $\alpha_\Phi$, then there exists $s'$ such that $s'$ is visited during the
  $\algname(M,\Phi)$ and $\alpha_\Phi(s) = \alpha_\Phi(s')$.
\end{lemma}

\begin{proof}
  By induction with respect to the number of exact
  transitions from the initial state necessary for reaching the state $s$.
  Basic step ($k=0$) is trivial. For the induction step, suppose that state $s$
  is reachable via sequence of exact transitions: $s_0 \myarrow{i_0} \ldots
  \myarrow{i_{k-1}} s_k \myarrow{i_k} s_{k+1} = s$. By the induction hypothesis there
  exists $s_k'$ such that $s_k'$ is visited and $\alpha_\Phi(s_k') =
  \alpha_\Phi(s_k)$.  Because the abstraction is exact with respect to $s_k
  \myarrow{i_k} s$, there must be $s'$ such that $s_k' \myarrow{i_k} s'$ and
  $\alpha_\Phi(s') = \alpha_\Phi(s)$. This successor $s'$ is considered during
  the visit of $s_k'$. There are two cases to be analyzed.
  \begin{enumerate}
  \item $s'$ is added to $|Wait|$ and later visited,
  \item $s'$ is matched to a previously visited state $s''$ such that
    $\alpha_\Phi(s')=\alpha_\Phi(s'')$.
  \end{enumerate}
  In both cases we get that some state with the same abstract counterpart as
  $s$ is  visited during the search.~$\Box$
\end{proof}

\begin{lemma}
  \label{lemma:must-search-concrete}
  $\RL(\algname(M,\Phi))\subseteq\RL(\semantics{M})$.
\end{lemma}

\begin{proof}
  It is easy to verify that the following is an invariant of the search:
  '$|Wait|$' is a subset of reachable states in $\semantics{M}$. The lemma follows.~$\Box$
\end{proof}

\begin{lemma}
  \label{lemma:bisim-ok}
  Let $AP \subseteq \Phi$.  If for all reachable states $s_1, s_2$ it holds that 
  $\alpha_\Phi(s_1) = \alpha_\Phi(s_2) \Rightarrow s_1 \sim s_2$, then
  $\algname(M, \Phi) \sim \semantics{M}$.
\end{lemma}

\begin{proof}
  Consider relation $R$ defined as: $s_1 R s_2$ iff $s_1=s_2$ or $s_1$
  is matched to $s_2$. Then $R$ is a bisimulation
  relation between $\algname(M, \Phi)$ and $\semantics{M}$.~$\Box$
\end{proof}

\begin{lemma}
  \label{lemma:check-correctness}
  Let $(A, \Phi_{new}) = \algname(M, \Phi)$.  If $\Phi_{new} = \Phi$,
  then \mbox{$A \sim \semantics{M}$}.
\end{lemma}

\begin{proof}
  Due to Lemma~\ref{lemma:bisim-ok} it is sufficient to show that if
  $\Phi_{new} = \Phi$ then $\alpha_\Phi$ induces a bisimulation relation on the
  reachable part of the transition system~$\semantics{M}$.  We first show that
  every reachable state in $\semantics{M}$ is reached by exact transitions. We
  proceed on induction by the number of transitions from the initial state to
  $s$. Basic step ($k=0$) is trivial. For the induction step, suppose that
  state $s$ is reachable via a sequence of exact transitions of length $k$. By
  Lemma~\ref{lemma:must-sequences} some state $s'$ such that $\alpha_\Phi(s) =
  \alpha_\Phi(s')$ is visited during the search. During the visit of the state
  $s'$ we check exactness of the abstraction (see
  Section~\ref{sec:check-exact-abstr}). Since $\Phi_{new} = \Phi$ it follows
  that the abstraction is exact for $s'$, i.e., $s' \notarrow{i}$ iff $s
  \notarrow{i}$ and for every outgoing transition 
  $s' \myarrow{i} s_1'$ and $\alpha(s)=\alpha(s')$ there exists $s_1$ such that
  $s\myarrow{i}s_1$ and $\alpha(s_1)=\alpha(s'_1)$.  Since $i$ is
  deterministic, it follows that $s_1$ is the only successor of $s$ by
  transition $i$ and transition $s \myarrow{i} s_1$ is also exact. Moreover, it
  satisfies the same criterion for bisimulation, i.e. for all $s''$ such that
  $\alpha_\Phi(s) = \alpha_\Phi(s'')$ there exists $s_1''$ such that
  $\alpha_\Phi(s_1) = \alpha_\Phi(s_1'')$ and $s'' \myarrow{i} s_1''$.~$\Box$ 
\end{proof}

\begin{proof}[of Theorem~\ref{theorem:correctness}]
  The first claim follows from the fact that $\algname$ produces
  an under-approximation (Lemma~\ref{lemma:must-search-concrete}). The second
  claim follows from Lemma~\ref{lemma:check-correctness}.~$\Box$
\end{proof}

\subsubsection*{Proof of Theorem~2}

In order to prove Theorem~2, we study sequences $\{ A_j \}_{j=0}^\infty$ of transition systems 
generated during $\refalgname$. We assume that $\algname$ uses breadth-first search order.
The basic idea of the proof is that any two states that are in different
bisimulation classes ($s \not \sim s'$) are eventually distinguished by the
abstraction function, i.e. $\exists j$ such that $\alpha_{\Phi_j}(s)
\neq\alpha_{\Phi_j}(s')$ (Lemma~\ref{lemma:states-distinguieshed}). 
Moreover, each bisimulation class of $\semantics{M}$ is eventually
visited by $\refalgname$ (Lemma~\ref{lemma:each-bisim-class-reached}) 
and the finite set of reachable labelings emerges (Lemmas~\ref{lemma:all-labelings}
and~\ref{lemma:finite-quotient}).

\begin{lemma}
  \label{lemma:states-distinguieshed}
  Let $\{ A_j \}_{j=0}^\infty$ be a sequence of transition systems generated
  during an infinite run of $\refalgname$ and $\infm = \{ s \mid $ there exists
  infinitely many $j$ such that $s \in A_j \}$.  If $s \not\sim s'$ and $s \in
  \infm$ then there exists $j$ such that $\alpha_{\Phi_k}(s) \neq
  \alpha_{\Phi_k}(s')$ for all $k \geq j$.
\end{lemma}

\begin{proof}
  By induction with respect to $k$ where $k$ is the smallest
  number such that $s \not \sim_k s'$.  Basic step: for $k=0$ it means that
  $L(s) \neq L(s')$ and therefore $\alpha_{\Phi_1}(s) \neq
  \alpha_{\Phi_1}(s')$. Induction step $(k+1)$: Let $s_1, s_1'$ be such that $s
  \myarrow{i} s_1, s' \myarrow{i} s_1'$ and $s_1 \not \sim_k s_1'$. Since $s$
  is visited in infinitely many iterations of $\algname$, $s_1$ is considered
  in infinitely many iteration of $\algname$ and therefore one of the following
  must hold:

  \begin{enumerate}
  \item State $s_1 \in \infm$. Then we can apply induction hypothesis, i.e. 
    there exits $j$ such that $\alpha_{\Phi_k}(s_1) \neq \alpha_{\Phi_k}(s_1')$ for all $k\geq j$.
    
  \item State $s_1$ is matched to some state in infinitely many runs of
    $\algname$. Since we use breadth-first order, there are only finitely many
    states to which it can be matched (with breadth-first search order the
    state can be matched only to states with lower or equal distance from the
    initial state). Therefore, there exists a state $s_2$ such that $s_1$ is
    matched to $s_2$ in infinitely many runs of $\algname$, this means that
    $\alpha_{\Phi_j}(s_1) \neq \alpha_{\Phi_j}(s_2)$ for all $j$. From the induction
    hypothesis we get that $s_1 \sim_k s_2$ and hence $s_2 \not\sim_k s_1'$.
     Moreover, from the induction hypothesis we get that there exists $m$ such that
    $\alpha_{\Phi_k}(s_2) \neq \alpha_{\Phi_k}(s_1')$ for all $k\geq m$. Therefore
    $\alpha_{\Phi_k}(s_1) \neq \alpha_{\Phi_k}(s_1')$ for all $k\geq m$.
  \end{enumerate}
  
 \noindent In both cases we get that there exists $j$ such that $\alpha_{\Phi_j}$ is not
  exact with respect to $s \myarrow{i} s_1$, therefore
  $wp(\alpha_{\Phi_j}(s_1), t_i)$ will be included in $\Phi_{j+1}$ and
  therefore $\alpha_{\Phi_{j+1}}(s_1) \neq \alpha_{\Phi_{j+1}}(s_1')$.~$\Box$
\end{proof}

\begin{lemma}
  \label{lemma:each-bisim-class-reached}
  For each reachable bisimulation class $B$ of $\semantics{M}$ there exists a
  state $s \in B$ such that $s$ is visited by $\refalgname(M, \varphi)$
  infinitely often.
\end{lemma}

\begin{proof}
  By induction with respect to the length of the shortest path by
  which some state from $B$ is reachable.  Basic step is obvious. Induction step:
  let state from $B$ be reachable via path $s_0 \myarrow{i_0} \ldots
\myarrow{i_{k-1}} s_k \myarrow{i_k} s_{k+1}$. By
  induction hypothesis some state $s' \sim s_k$ is reached during the
refinement search infinitely often. Consider state $s''$ such that $s'
\myarrow{i_k} s''$. It holds that $s'' \sim s_{k+1}$ and from
Lemma~\ref{lemma:states-distinguieshed} we get that $s''$ is visited infinitely
often.~$\Box$
\end{proof}

\begin{lemma}
\label{lemma:all-labelings}
Let $\{ A_j \}_{j=0}^\infty$ be a sequence of transition systems generated
during an infinite run of $\refalgname(M,\varphi)$.  There exists $j$ such that $RL(A_j) =
RL(\semantics{M})$.
\end{lemma}

\begin{proof}
  For each $l \in RL(\semantics{M})$ we choose some bisimulation class $B$ such
  that $s\in B \Rightarrow L(s) = l$. In this way we obtain a finite set of
  bisimulation classes $\{B_1, \ldots, B_k\}$ which covers all labels in
  $RL(\semantics{M})$. Note that $RL(\semantics{M})$ is finite because $AP$ is
  finite. Now we show that there exists an iteration in which at least one
  state from each of these classes is visited. This is done similarly to the
  proof of Lemma~\ref{lemma:each-bisim-class-reached}.~$\Box$
\end{proof}

\begin{lemma}
  \label{lemma:finite-quotient}
  Let $\{ A_j \}_{j=0}^\infty$ be a sequence of transition systems generated
  during an infinite run of $\refalgname$.  If the reachable part of
  the bisimulation quotient is finite, then there exists $j$ such that $A_j \sim
  \semantics{M}$.
\end{lemma}

\begin{proof}
  By contradiction. Suppose that $\forall j: A_j \not \sim \semantics{M}$. From
  Lemma~\ref{lemma:bisim-ok} we get that there exists reachable $s, s'$ such
  that $\forall j: \alpha_{\phi_j}(s) = \alpha_{\phi_j}(s')$ and $s \not \sim
  s'$. We show (similarly to the proof of Lemma~\ref{lemma:must-sequences})
  that there exists such $s$ which is visited infinitely often.  From
  Lemma~\ref{lemma:states-distinguieshed} we get that eventually
  $\alpha_{\phi_j}(s) \neq \alpha_{\phi_j}(s')$ which is the contradiction.~$\Box$
\end{proof}

\begin{proof}[of Theorem~\ref{th:termination}]
  This theorem is a direct consequence of
  Lemmas~\ref{lemma:all-labelings} and~\ref{lemma:finite-quotient}.~$\Box$ 
\end{proof}

\section{Properties}
\label{sec:discussions}

Having discussed correctness and termination, we now turn to
other interesting properties of the algorithm. 

\subsection{Non-termination for Finite State System}
\label{sec:non-termination}
We should note that the proposed iterative algorithm is not guaranteed to
terminate even for a finite state program. This situation is illustrated by the
example from Figure~\ref{fig:incomplete-refinement}; $x$ and $y$ are initialized to zero. 
The property that we check is that "pc=1" 
is unreachable. Although the program is
finite state (and therefore the problem can be easily solved with classical
explicit model checking), it is quite difficult to solve using abstraction
refinement techniques.  The iterative algorithm does not terminate on this
example: it  keeps adding predicates $y\geq 0, y+x \geq 0, y+2x\geq 0,
\ldots$. Note that, in accordance with Theorem~\ref{th:termination}, it 
eventually produces a bisimilar structure.  However, the algorithm is not
able to detect termination, and it  keeps refining indefinitely. The reason
is that the algorithm keeps adding predicates that refine the unreachable part
of the system under analysis.

Also note that the same problem occurs with over-approximation based
abstraction techniques that use refinement based on weakest precondition
calculations~\cite{MAGIC,734102}. Those techniques introduce the same
predicates. Moreover, unlike our technique, they will keep generating {\em spurious} counter-examples. 
For this example no may/must abstraction based on predicates and refinement with weakest 
precondition calculations can produce a structure that is bisimilar to the concrete system 
(the concrete system is rather trivial --- it has only one state).

This example also illustrates another difference between the method presented here and over-approximation based 
predicate abstraction with refinement, in particular~\cite{734102}. If the analyzed system has a 
{\em reachable} finite bisimulation quotient then our algorithm is guaranteed to find it 
(see Theorem~\ref{th:termination} and Lemma~\ref{lemma:finite-quotient}). 
In contrast, the method in~\cite{734102} will fail to compute a finite state abstraction for the example; 
this result seems to contradict the bisimulation completeness claim (Theorem~3) from~\cite{734102}. 
We conjecture that the method in~\cite{734102} is not guaranteed to compute a finite state abstraction unless 
both the {\em reachable and unreachable} quotient is finite. 

  \begin{figure}
    \begin{center}
      \[ \begin{array}{lll}
         pc = 0 \wedge y \geq 0 & \longmapsto & y := y + x\\
         pc = 0 \wedge y < 0 & \longmapsto & pc := 1\\
      \end{array} \]
      \caption{Example illustrating non-terminating refinement for finite state systems}
      \label{fig:incomplete-refinement}
    \end{center}
  \end{figure}

To solve the problem of non-termination for finite state systems, we propose to
use the following heuristic. If there is a transition for which we cannot prove
that the abstraction is exact in several subsequent iterations of the
algorithm, then we add predicates describing the concrete state; i.e. in the
example from Figure~\ref{fig:incomplete-refinement} we would add predicates
$x=0$ and $y=0$. The abstraction eventually becomes exact with respect to each
transition. And since the number of reachable transitions is finite, the
algorithm must terminate.

\begin{corollary}
  If the reachable part of $\semantics{M}$ is finite state then the modified algorithm terminates.
\end{corollary}

\subsection{Search Order and Non-Monotonicity}
The search order used in $\algname$ (depth-first or breadth-first) influences 
the size of the generated structure, the newly computed predicates, and even 
the number of iterations of the main algorithm. If there are two states $s_1$ 
and $s_2$ such that $\alpha_\Phi(s_1) = \alpha_\Phi(s_2)$ but 
$s_1 \not \sim s_2$ then, depending on whether $s_1$ or $s_2$ is visited 
first, different parts of the transition system will be explored. For our implementation,
we use breadth-first search order.

\begin{figure}
  \begin{center}

    \begin{tabular}{ccc}
      Program & &State space \\
      & &\\
      \includegraphics[scale=0.5]{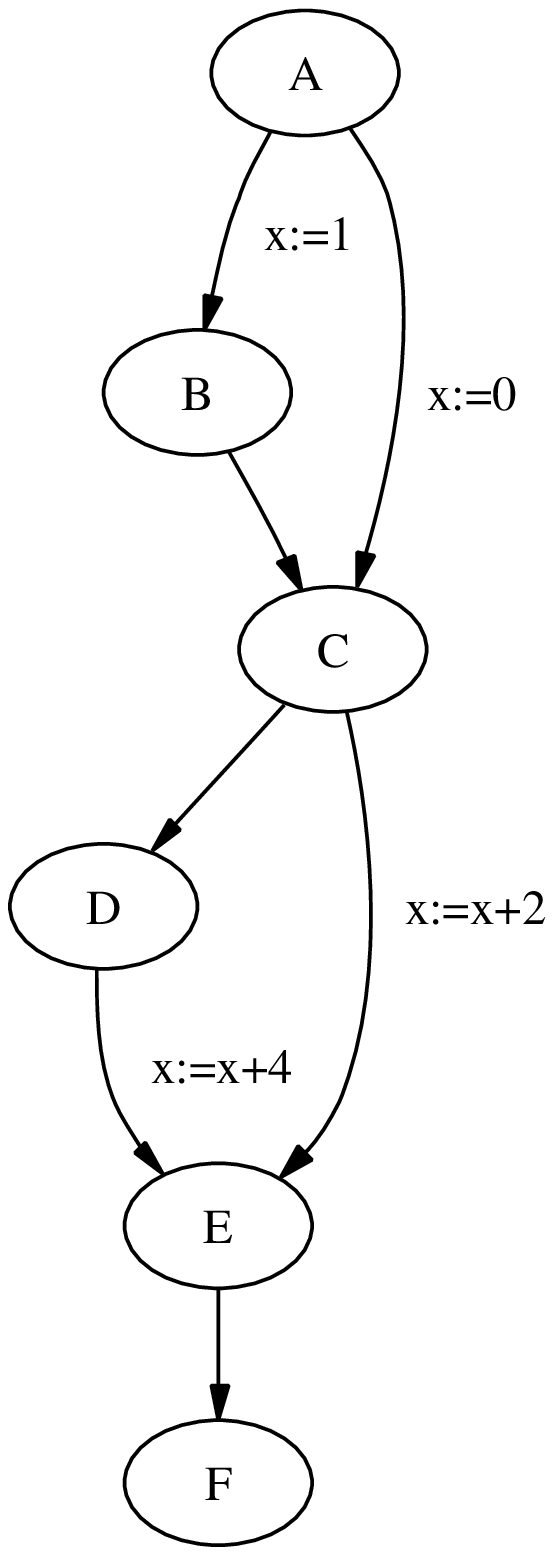} & \ \ \ \ \ \ &
      \includegraphics[scale=0.5]{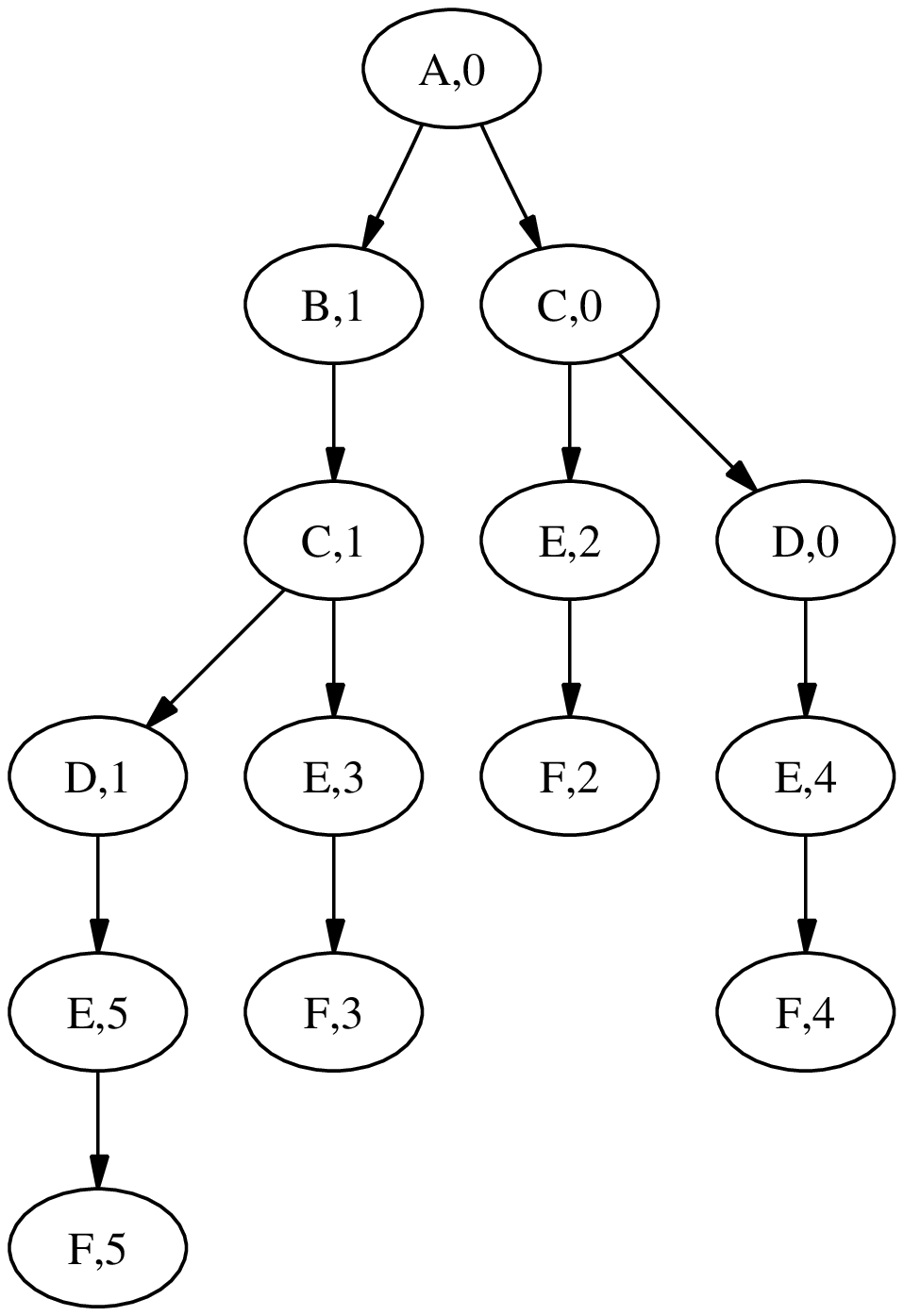} \\
    \end{tabular}
    \caption{Example illustrating non-monotonic refinement}
    \label{fig:example-non-monotonne}
  \end{center}
\end{figure}

Also note that the refinement algorithm is non-monotone, i.e. a labeling
which is reachable in one iteration may not be reachable in the next iteration.
However, the algorithm is guaranteed to converge to the correct answer.
The example in Figure~\ref{fig:example-non-monotonne} illustrates this
non-monotonic behavior. Figure~\ref{fig:example-non-monotonne} (left) shows the
transitions of the example program (for clarity of presentation, we depict the
program in a graphical notation); the program has only one variable~$x$; the
program counter ranges over $A$, $B$, $C$ \ldots.
Figure~\ref{fig:example-non-monotonne} (right) shows the whole concrete state
space of the program. As usual, states are labeled by the program counter and
the concrete value of program variable~$x$.  Let us consider the first
iteration of the algorithm, with abstraction predicate $x\geq 3$ and with
breadth-first search order -- the following states are visited: $(A,0)$,
$(B,1)$, $(C,0)$, $(E,2)$, $(D,0)$, $(F,2), (E,4), (F,4)$.  Assume now that the
refinement step adds a new predicate $x=1$; then, in the second iteration, the
following states are visited: $(A,0), (B,1), (C,0)$, $(C,1)$, $(E,2)$, $(D,0)$,
$(D,1)$, $(E,3)$, $(F,2)$, $(F,3)$.  States $(E,4)$ and $(F,4)$ are visited
during the first iteration and they are not visited during the second one.

\subsection{Relation to Other Abstractions}
\label{sec:other-abstractions}

We discuss now the relationship between our abstraction based iterative 
algorithm and other (under-approximating)
abstractions, in particular with the {\em must} abstractions 
from~\cite{ShohamG03,ShohamG04} and with the abstractions induced by the refined definition of 
{\em must} transitions presented in~\cite{BallKupfermanyorshCAV05}.
We first remark that the abstract state space explored by our approach is (potentially) a better 
approximation than the {\em must} abstraction. This is formulated by the following lemma. 
\begin{lemma}  
  Let $AP \subseteq \Phi$. Then $\RL(\algname(M,\Phi))$ is a superset of the
  reachable labelings in the {\em must} abstraction induced by $\Phi$.
\end{lemma}

\begin{proof}
  The lemma is a direct consequence of Lemma~\ref{lemma:must-sequences}.~$\Box$
\end{proof}

As mentioned, the iterative refinement in our algorithm is non-monotonic. A similar problem 
occurs in the context of {\em must} abstractions: the set of
{\em must} transitions is not generally monotonically non-decreasing when
predicates are added to refine an abstract system~\cite{701647,ShohamG04}. This problem is 
addressed in~\cite{ShohamG03,ShohamG04}, by creating {\em hyper must} transitions (representing 
sets of {\em must} transitions). Note that the approaches presented 
in~\cite{BallKupfermanyorshCAV05,ShohamG03,ShohamG04} require the a-priori construction of 
abstract {\em must} (and {\em hyper must}) transitions and therefore make an exponential number 
of theorem prover calls. In contrast our approach does not require the computation of abstract 
transitions, since it executes directly the concrete transitions (and it only makes theorem prover 
calls during refinement).

Recently, Ball et al.~\cite{BallKupfermanyorshCAV05} defined an extension of the {\em must}
abstraction based on so called $\mustaa^-$ transitions: $a_1 \mustminus{i} a_2$
iff for all concrete states $s_2$ such that $\alpha_\Phi(s_2)=a_2$, there 
exists concrete state $s_1$ such that $\alpha_\Phi(s_1)=a_1$ and 
$s_1\myarrow{i} s_2$ (for some $\Phi$).

They call the classical {\em must} transitions $\mustaa^+$ transitions and they describe a 
reachability analysis that uses both $\mustaa^-$ and $\mustaa^+$ transitions; the set of 
reachable labelings is
defined as $ \{ L(s) \mid s\in S: s_0 \mustminus{}^* s_i \mustplus{}^* s \}$. 
This results in an under-approximation of the set $\RL(\semantics{M})$ and at
the same time it is a better under-approximation then the one obtained by
classical {\em must} transitions.

Here we show that under-approximations based on $\mustaa^+$/$\mustaa^-$
transitions and our algorithm based on $\algname$ are incomparable. The (trivial) example in
Figure~\ref{fig:example-comparison-must-alpha}~(a) illustrates that $\algname$
can be more precise than the analysis based on $\mustaa^+$/$\mustaa^-$
transitions.  If we consider the abstraction with respect to a single predicate
$x \geq 0$ we see that the program transition is neither $\mustaa^+$ nor
$\mustaa^-$ (hence the set of reachable labelings produced by the analysis
from~\cite{BallKupfermanyorshCAV05} contains only a labeling $x\geq 0$) whereas
$\algname$ executes the transition and finds a labeling $x<0$.

On the other hand, consider the example in
Figure~\ref{fig:example-comparison-must-alpha}~(b) and an abstraction with
respect to a single predicate $x \geq 3$. Due to state matching on the states
represented by $(pc=1,x=1)$ and $(pc=1,x=2)$, $\algname$  computes a
different set of labelings, depending on which of the first two transitions is
traversed first from the initial state. Therefore, the resulting set of
reachable labelings contains only one of $(pc = 3, x < 3)$, $(pc=3, x\geq 3)$.
Under-approximation based on $\mustaa^+$/$\mustaa^-$ transitions contains
both of these labelings.

\begin{figure}[tb]
  \begin{center}

    \begin{tabular}{ll}
      (a)\ \ \  & $ \begin{array}{lll}
          x \geq 0  &  \longmapsto & x := x - 1  \\
      \end{array} $  \\[6mm]
      (b) & $ \begin{array}{lll}
          \mathit{pc} = 0  &  \longmapsto & \mathit{pc} := 1, x := 1 \\        
          \mathit{pc} = 0  &  \longmapsto & \mathit{pc} := 1, x := 2 \\
          \mathit{pc} = 1  &  \longmapsto & \mathit{pc} := 2, x := x + 1 \\
          \mathit{pc} = 2 \wedge x \geq 3 &  \longmapsto & \mathit{pc} := 3\\
          \mathit{pc} = 2 \wedge x < 3 &  \longmapsto & \mathit{pc} := 3 \\
      \end{array} $ \\
    \end{tabular}

    \caption{Examples showing that under-approximations based on
      $\algname$  and $\mustaa^+$/$\mustaa^-$ transitions are incomparable}
    \label{fig:example-comparison-must-alpha}
  \end{center}
\end{figure}
\vfill\eject

\section{Extensions}
\label{sec:extensions}

In this section we propose several extensions of the main algorithm.

\subsection{Open Systems}
Until now, we have discussed our approach in the context of ``closed'' systems. However,
the approach can be extended to handling ``open'' systems (i.e. programs with inputs).
In order to model open systems, we extend the guarded commands language by allowing
assignments of the form $x := \mathit{input}$, which assigns to program variable $x$
an arbitrary value from the input domain (in our case the set of integers). 
We can also allow the initial values of the program variables to be unspecified,
in which case the transition system representing the open program has several 
(possibly unbounded) initial states.

In order to apply our approach, we need to compute, for each input variable,  
explicit concrete values that drive the concrete execution of the program.
What we really want here is to pick one input value for each satisfiable
valuation of the abstraction predicates. 
We can directly use the original algorithm --- it will simply try all the
possible values and continue the program execution only from values that satisfy the predicate
combinations (most of the states that contain such input values will be matched if
they lead to the same valuation of abstraction predicates). 
This ``brute force'' approach requires enumerating eventually the whole input domain, 
which is impossible for infinite input domains. Note however that the approach might 
still be very useful at detecting errors.

Alternatively, we can use a constraint solver for computing the input values that are solutions of 
the satisfiable combinations of abstraction predicates (provided that satisfiability is 
decidable for the abstraction predicates). 
The decision whether to use the ``brute force'' approach or the satisfiability approach depends 
on the number of abstraction predicates and the size of the input domain. With the brute force 
approach, the the whole input domain needs to be enumerated eventually. With the satisfiability 
approach, there are at most $2^k$ satisfiability queries (where $k$ is number of predicates 
which depend on the input variable).

\subsection{Transition Dependent Predicates}

The predicates that are generated after the validity check for one
transition are used `globally' at the next iteration. This may
cause unnecessary  refinement --- the new predicates may
distinguish states which do not need to be distinguished. To avoid
this, we could use `transition dependent' predicates. The idea is to
associate the abstraction predicates with the program counter
corresponding to the transition that generated them. New
predicates are then added only to the set of the respective
program counter. However, with this approach,
it may take longer before predicates are
`propagated' to all the locations where they are needed, i.e. more
iterations are needed before an error is detected or an exact
abstraction is found. We need to further investigate these issues.
Similar ideas are presented 
in~\cite{ChakiClarkeGroceStrichmanCHARME03,HenzingerJhalaMajumdarMcMillanPOPL04}, 
in the context of over-approximation based predicate abstraction.

\subsection{Light-weight Approach}

As mentioned, the under-approximation and refinement approach can be used in a
lightweight but systematic manner, without using a theorem prover for validity checking. 
Specifically, for each
explored transition $t_i$ 
refinement adds the new predicates from
$\alpha_\Phi(s')[e_i(\vec{x})/\vec{x}]$, regardless of the fact that the
abstraction is exact with respect to transition $t_i$. This
approach may result in unnecessary refinement. A similar
refinement procedure was used in \cite{734102} for over-approximation predicate abstraction.

We are also considering several heuristics for
generating new abstraction predicates. For example, it is
customary to add the predicates that appear in the guards and in
the property to be checked. One could also add predicates
generated dynamically, using tools like Daikon~\cite{Daikon}, or
predicates from known invariants of the system, generated using
static analysis techniques. Section~\ref{sec:applications} shows an example where a 
statically computed invariant helped with the termination of the presented iterative algorithm.

In order to extend the applicability of the proposed technique to the analysis
of full-fledged programming languages, we are investigating abstractions that
record information about the shape of the program heap, to be used in
conjunction with the abstraction predicates. We have reported about these
experiments in~\cite{ase05testing}.

\section{Implementation and Applications}
\label{sec:applications}

We have implemented our approach for the guarded command language. Our
implementation is done in the language
Ocaml\footnote{\texttt{http://caml.inria.fr/}} and it uses the Simplify theorem
prover~\cite{Simplify}. The
implementation has just 590 lines of code (parsing + definition of semantics:
390 lines, $\algname$ algorithm: 170 lines, $\refalgname$ algorithm:
30 lines).
The implementation uses several optimizations for
reducing the number of theorem prover calls:

\begin{itemize}
\item When updating $\Phi_{new}$ for refinement, we add only those conjuncts of
  $\alpha_\Phi(s')[e_i(\vec{x})/\vec{x}]$ for which we cannot prove validity.
\item We cache queries to ensure that Simplify is not called
  twice for the same query.
\item All queries have the form of implication. Before calling the theorem prover
  for the implication, we check whether the right hand side is a tautology (in
  such case the implication is clearly satisfied). The results of these checks are
  also cached. 
\end{itemize}

\subsection{Experiments}
We discuss the application of our implementation for error detection and property verification in 
several multi process programs. The examples are:
the ticket mutual exclusion protocol, RAX (Remote Agent Experiment), 
a component extracted from an embedded spacecraft-control application, and the bakery mutual 
exclusion protocol. We also analyzed a single process device driver taken from~\cite{380932}, 
which is a ``classic'' example analyzed with predicate abstraction techniques. 
We analyzed defective and correct versions of each example program.
The RAX and device driver had known errors that we checked for. For the other examples, 
we seeded faults to obtain the defective versions.

Note that in the described experiments, we always start the first iteration of
the refinement algorithm with the program predicates which occur in guards.
All the reported results are for the breadth-first search order.

\begin{table}
   \begin{center}
    \begin{tabular}{|c||c|c|c|c|c|}
      \hline
      Example & Iterations & Concrete States & Abstract States & New Predicates & Queries  \\
              &            & (per iteration) & (per iteration) &                &          \\
      \hline
      ticket2-err  & 2 & 15, 31  & 9, 17 & 5  & 38   \\
      ticket3-err  & 1 & 102    & 44   & 4  & 14   \\
      RAX-err      & 1 & 69     & 44   & 0  & 10   \\
      bakery-err   & 1 & 356    & 191  & 14 & 89  \\
      driver-err   & 1 & 10     & 10   & 0  & 2  \\
      \hline
      ticket2  & 4  & 15, 15, 15, 15    & 9, 9, 9, 9        & 6  &  124   \\
      ticket3  & 5  & 52, 58, 58, 58, 58 & 25, 31, 31, 31, 31 & 11 &  603  \\
      RAX      & -- & -- & --   & --             & --   \\
      bakery   & 3  & 278, 410, 537    & 152, 221, 292    & 24 & 1598  \\
      driver   & 2  & 10             & 9              & 0  & 7      \\
      \hline
    \end{tabular}
    
     \caption{Experimental results}
     \label{table:results-general}
   \end{center}
 \end{table}

Table~\ref{table:results-general} summarizes the results for each of the runs 
of our algorithm. The first part of the table  reports the analysis results 
for the defective examples (denoted with the {\tt -err} suffix), while the 
second part of the table reports the results for the correct examples.
For each example we report numbers for: refinement iterations, generated 
concrete states and stored abstract states,  generated predicates, and  
queries to the theorem prover. A "--" for RAX denotes that our analysis did 
not finish for this example (see discussion below).
Note that for the concrete and abstract states, we report {\em separate} 
numbers {\em for each iteration}. 
For example, running our tool on the error version of the ticket protocol 
with two processes ({\tt ticket2-err}) discovered the error after 2 
iterations; in the first iteration, the tool generated 15 concrete states and 
it stored 9 abstract states, while in the second  iteration, it generated 31 
concrete states and it stored 17 abstract states. 
We discuss the experiments in more detail below (full details are available 
at~\cite{radek-thesis}).

\subsection{Ticket Protocol}

This is a protocol for mutual exclusion~\cite{andrews-cp};
we use the formalization of the algorithm from~\cite{BultanGP97}. The
algorithm is based on a simple ``ticket'' procedure: a process which wants to
enter the critical section draws a ticket number that is one larger than the number held
by any other process. The process then waits until all processes with smaller
numbers are served: this is checked by a ``display'' variable which shows the
value of the ticket number which is currently the smallest. The model of the protocol is
given in Figure~\ref{fig:ticket-example}. The property of interest is mutual
exclusion in critical section ($\neg (\mathit{pc}_1 = 2 \wedge \mathit{pc}_2 =
2 \vee \mathit{pc}_2 = 2 \wedge \mathit{pc}_3 = 2 \vee \mathit{pc}_1 = 2 \wedge
\mathit{pc}_3 = 2)$). The state space is infinite (the ticket numbers increase
without any bound), but it has a finite bisimulation quotient.

\begin{figure}
  \begin{center}
    \[
    \begin{array}{lll}
       \mathit{pc}_1 = 0 & \longmapsto &  \mathit{pc}_1 := 1, \mathit{a}_1 := t, t := t + 1  \\ 
       \mathit{pc}_1 = 1 \wedge \mathit{a}_1 \leq s & \longmapsto &  \mathit{pc}_1 := 2  \\ 
       \mathit{pc}_1 = 2 & \longmapsto &  \mathit{pc}_1 := 0, s := s + 1  \\[2mm]
       \mathit{pc}_2 = 0 & \longmapsto &  \mathit{pc}_2 := 1, \mathit{a}_2 := t, t := t + 1  \\ 
       \mathit{pc}_2 = 1 \wedge \mathit{a}_2 \leq s & \longmapsto &  \mathit{pc}_2 := 2  \\ 
       \mathit{pc}_2 = 2 & \longmapsto &  \mathit{pc}_2 := 0, s := s + 1  \\[2mm]
       \mathit{pc}_3 = 0 & \longmapsto &  \mathit{pc}_3 := 1, \mathit{a}_3 := t, t := t + 1  \\ 
       \mathit{pc}_3 = 1 \wedge \mathit{a}_3 \leq s & \longmapsto &  \mathit{pc}_3 := 2  \\ 
       \mathit{pc}_3 = 2 & \longmapsto &  \mathit{pc}_3 := 0, s := s + 1  \\ 
    \end{array}
    \]

    \caption{Ticket protocol (instance for three processes)}
    \label{fig:ticket-example}
  \end{center}
\end{figure}

We used our tool to prove successfully that the property holds. We analyzed
several versions of the protocol. The intermediate analysis results for the 
protocol with three processes are given
in Table~\ref{table:ticket-example-results}. We report the
following results for each iteration of the refinement algorithm: the number of
generated concrete states, the number of stored abstract states, the number of
queries to the theorem prover, the number of hits to a queries cache, and the
newly generated predicates.

\begin{table}
   \begin{center}
    \begin{tabular}{|c|c|c|c|c|c|}
      \hline
      Iteration & Concrete & Abstract & Num.  & Cache  & New predicates \\
       & states & states & queries & hits &  \\
      \hline
      1 & 52  & 25 & 14 & 18 & $a_1 \leq s+1, a_2 \leq s+1, a_3 \leq s+1,$ \\
      & & & & & $t \leq s $ \\
      2 & 58 & 31 & 70 & 152 & $a_1 \leq s+2, a_2 \leq s+2, a_3 \leq s+2,$ \\     
      & & & & & $ t \leq s +1, t+1 \leq s $ \\
      3 & 58 & 31 & 151 & 475 & $t \leq s+2$ \\
      4 & 58 & 31 & 173 & 585 & $t \leq s+3 $ \\
      5 & 58 & 31 & 195 & 657 & - \\
      \hline
    \end{tabular}
    
     \caption{Ticket protocol  for three processes: intermediate results}
     \label{table:ticket-example-results}
   \end{center}
 \end{table}

 As discussed, we also seeded an error in the protocol and used our tool for 
error detection. The error was seeded by changing the assignment $s := s + 1$ 
into $s := s + 2$. For an instance with two processes the error is found after 
two iterations. For an instance with three processes the error state can be
reached by suitable interleaving in the first round of the protocol and the
tool finds the error in the first iteration. 

\subsection{RAX}

The RAX example (illustrated in Figure~\ref{fig:rax-example}) is derived from
the software used in the NASA Deep Space 1 Remote Agent experiment, which
deadlocked during flight~\cite{VisserParkPenix00}. We encoded the deadlock
check as ``$pc_1=4 \wedge pc_2=5\wedge w_1=1 \wedge w_2=1$ is unreachable''.
The error is found after one iteration; the reported counter-example has 8
steps.

Note that the state space of the program is unbounded, as the program keeps
incrementing the counters $e_1$ and $e_2$, when $pc_2=2$ and $pc_1=6$,
respectively. We also ran our algorithm to see if it converges to a finite
bisimulation quotient.  Interestingly, the algorithm does not terminate for the
RAX example, although it has a finite reachable bisimulation quotient. The 
results are shown in Table~\ref{table:rax-example-results}. However, if
we assume that the counters in the program are non-negative, i.e. we introduce
two new predicates, $e1 \ge 0$, $e2 \ge 0$ (which can be easily discovered 
using static analysis), then the algorithm terminates after two iterations. 
The tool reports the following results : 69 concrete and 
44 abstract states explored in the first iteration, 101 concrete and 65 
abstract states in the second iteration, two new predicates and 40 queries.

\begin{figure}
  \begin{center}
    \[ 
    \begin{array}{lll}
       \mathit{pc}_1 = 1 & \longmapsto &  \mathit{c}_1 := 0, \mathit{pc}_1 := 2  \\ 
       \mathit{pc}_1 = 2 \wedge \mathit{c}_1 = \mathit{e}_1 & \longmapsto &  \mathit{pc}_1 := 3  \\ 
       \mathit{pc}_1 = 3 & \longmapsto &  \mathit{w}_1 := 1, \mathit{pc}_1 := 4  \\ 
       \mathit{pc}_1 = 4 \wedge \mathit{w}_1 = 0 & \longmapsto &  \mathit{pc}_1 := 5  \\ 
       \mathit{pc}_1 = 2 \wedge \mathit{c}_1 \neq \mathit{e}_1 & \longmapsto &  \mathit{pc}_1 := 5  \\ 
       \mathit{pc}_1 = 5 & \longmapsto &  \mathit{c}_1 := \mathit{e}_1, \mathit{pc}_1 := 6  \\ 
       \mathit{pc}_1 = 6 & \longmapsto &  \mathit{e}_2 := \mathit{e}_2 + 1,
      \mathit{w}_2 := 0, \mathit{pc}_1 := 2  \\[2mm]
       \mathit{pc}_2 = 1 & \longmapsto &  \mathit{c}_2 := 0, \mathit{pc}_2 :=2  \\ 
       \mathit{pc}_2 = 2 & \longmapsto &  \mathit{e}_1 := \mathit{e}_1 + 1, \mathit{w}_1 :=0 , \mathit{pc}_2 :=3  \\ 
       \mathit{pc}_2 = 3 \wedge \mathit{c}_2 = \mathit{e}_2 & \longmapsto &  \mathit{pc}_2 :=4  \\ 
       \mathit{pc}_2 = 4 & \longmapsto &  \mathit{w}_2 := 1, \mathit{pc}_2 :=5  \\ 
       \mathit{pc}_2 = 5 \wedge \mathit{w}_2 = 0 & \longmapsto &  \mathit{pc}_2 :=6  \\ 
       \mathit{pc}_2 = 3 \wedge \mathit{c}_2 \neq \mathit{e}_2 & \longmapsto &  \mathit{pc}_2 :=6  \\ 
       \mathit{pc}_2 = 6 & \longmapsto &  \mathit{c}_2 := \mathit{e}_2, \mathit{pc}_2 :=2  \\ 
    \end{array}
    \]
    \caption{RAX example}
    \label{fig:rax-example}
  \end{center}
\end{figure}

\begin{table}
  \begin{center}

    \begin{tabular}{|c|c|c|c|c|c|}
      \hline
      Iteration & Concrete & Abstract & Num.  & Cache  & New predicates \\
       & states & states & queries & hits &  \\
      \hline
       1 & 69 & 44 & 10 & 10 & $e_1 = 0, e_2 = 0$ \\
       2 & 101 & 65 & 20 & 44 & $e_1 = -1, e_2 = -1$ \\
       3 & 101 & 65 & 26 & 64 & $e_1 = -2, e_2 = -2$ \\
       4 & 101 & 65 & 32 & 84 & $\ldots$ \\
      \hline
    \end{tabular}
    
    \caption{RAX example: intermediate results}
    \label{table:rax-example-results}
  \end{center}
\end{table}

\subsection{Bakery Protocol}
This is another well-known protocol for mutual
exclusion. The protocol is similar to the ticket protocol (the ticket
protocol requires special hardware instruction like Fetch-and-Add, whereas
the bakery protocol is applicable without any special instructions). The
model has 10 variables. The property of interest is again mutual exclusion.
The state space is infinite with a finite bisimulation quotient. 
The property can be proved by the algorithm in three iterations, using 31 
predicates.
For this example, we seeded an error by changing a guard $\mathit{num}_1 <
\mathit{num}_0$ into $\mathit{num}_1 > \mathit{num}_0$ which creates a
nontrivial error in the protocol. The tool can find the error in the first
iteration.

\subsection{Device Driver} 
  This is a ``classic'' example analyzed using predicate
  abstraction~\cite{380932}.  The property of interest is the correct use of a
  lock. Our tool can prove that the property holds after one iteration (using
  just the predicates from guards): the algorithm explores 10 concrete states,
  9 abstract states and casts 3 queries to the theorem prover. For an
  erroneous version of the driver, the tool finds an error in the first
  iteration as well.

\subsection{Discussion}

These preliminary experiments show the merits of our approach. 
The approach proves to be effective in computing finite bisimilar 
structures of non-trivial infinite state systems and in finding errors using 
under-approximation based predicate abstraction. 
Of course, much more experimentation is necessary to really assess the 
practical benefits of the proposed technique and a lot more engineering is 
required to apply it to real programming languages. Extensions for handling 
complex features such as pointers, arrays and procedures, are tedious but 
conceptually not very hard.

We also note that in some cases (e.g. {\tt ticket2}, {\tt ticket3} and 
{\tt RAX}) the number of explored concrete and abstract states stays the 
same after the first iteration; however our algorithm needs more than two 
iterations to discover all the necessary abstraction predicates, according to 
the exactness criteria that we defined. The results suggest that it is 
possible to relax these criteria and still provide a guarantee that the 
relevant state space of the analyzed program has been explored. We leave this 
topic for future work.

\subsection{Comparison and Combination with Over-approximation Based Approaches}
\label{sec:over}
We should mention that the application of over-approximation based predicate abstraction to a 
Java version of RAX is described in detail in~\cite{VisserParkPenix00}. In that work, four 
different predicates were used to produce an abstract model that is bisimilar to the original 
program. In contrast, the work presented here allowed more aggressive abstraction to recover 
feasible counter-examples. Our technique explores transitions that are guaranteed to be feasible. 
In contrast, the over-approximation based techniques such as the ones 
from~\cite{759192,MAGIC,BLAST-IDEA} may also explore transitions that are spurious and 
therefore could require additional refinement before reporting a real counter-example. 

As mentioned, over-approximation based abstraction techniques involve exponentially many 
theorem prover queries (in the number of predicates), at each iteration. This computation is 
performed regardless of the size of (the reachable portion of) the analyzed system. In our case, 
theorem prover queries are only performed during refinement and they involve only the 
{\em reachable} state space of the system under analysis. On the other hand, over-approximation 
based techniques are good at proving properties (as they compute abstractions that are coarser 
than the bisimulation quotient but sufficient to prove safety properties). 
We believe however that the technique presented here is {\em complementary} 
to over-approximation abstractions and it should {\em combined} (rather than compared)
with such techniques. Our technique could be used for 
discovering efficiently feasible counter-examples in the space bounded by the abstraction 
predicates (that are used in the over-approximation analysis). In the future, we plan to study 
more the strengths and weaknesses of each approach and to investigate their {\em integration}.

\section{Related Work}
\label{sec:related-work}
Throughout the paper, we have already discussed the relationship between our work and 
predicate abstraction (see the previous section and also Section~\ref{sec:discussions}, where we 
compared our work with over-approximation approaches, in particular the work of Namjoshi and Kurshan~\cite{734102}, 
and with under-approximation approaches using {\em must} 
transitions~\cite{BallKupfermanyorshCAV05,ShohamG03,ShohamG04}).
We discuss here other approaches that are closely related to ours.

The work of Grumberg et al.~\cite{GRU05} uses a refinement of an
under-approximation to improve analysis of multi-process systems. The
procedure in \cite{GRU05} checks models with an increasing set of allowed
interleavings of the given processes, starting from a single interleaving. It
uses SAT-based bounded model checking for analysis and refinement, whereas here
we focus on explicit model checking and predicate abstraction, and we use
weakest precondition calculations for abstraction refinement.

Another closely related work is that of Lee and Yannakakis~\cite{LEE92}, which
proposes an on-the-fly algorithm for computing the bisimulation quotient of an
(infinite state) transition system.  Similar to our approach, the algorithm
from~\cite{LEE92} traverses concrete transitions while computing \emph{blocks}
of equivalent states; if some transition is found to be \emph{unstable} the
block is \emph{split} into sub-blocks. Note however that unlike~\cite{LEE92} 
our algorithm is geared towards error detection and it is formulated in terms 
of predicate abstraction with a clear separation between state exploration 
and refinement. There are other important differences between our approach and 
the work presented in~\cite{LEE92}. We use refinement globally while the block 
splitting in~\cite{LEE92} is local. This makes the approach in~\cite{LEE92} 
more efficient in the number of visited states. On the other hand, the global 
refinement has the advantage of faster propagating the new predicates across 
the system but it may lead to unnecessary refinement. As a consequence of 
this global refinement, our algorithm may not compute \emph{the} bisimulation 
quotient (as in~\cite{LEE92}) but rather just \emph{a} bisimilar structure 
(due to extra refinement). 
We view the experimental comparison of the two approaches as an interesting 
topic for future work. 

In previous work~\cite{choose-free}, we developed a technique for finding
feasible counter-examples in abstracted programs. The technique
essentially explores an under-approximation defined by the {\em must} abstract
transitions (although the presentation is not formalized in these terms). The
work presented here explores an under-approximation which is more precise than
the abstract system defined by the {\em must} transitions. Hence it has a
better chance of finding bugs while enabling more aggressive abstraction and
therefore more state space reduction.

Model-driven software verification~\cite{Holzmann04} advocates the use of
abstraction mappings during concrete model checking in a way similar to what we
present here. In their approach, the abstraction function needs to be provided
by the user.  The CMC model checking tool~\cite{CMC} also attempts to store
state information in memory using aggressive compressing techniques (which can
be seen as a form of abstraction), while the detailed state information is kept
on the stack. These techniques allow the detection of subtle bugs which can not
be discovered by classical model checking, using e.g. breadth first search  or
by state-less model checking~\cite{Verisoft}.  While these techniques use
abstractions in an ad-hoc manner, our work contributes the automated generation
and refinement of abstractions.

Directed automated random testing (DART)~\cite{DART} performs a concrete execution on random inputs
and it collects the {\em path constraints} along the executed paths. These path constraints are then
used to compute new inputs that drive the program along alternative paths.
The approach in~\cite{DART} is similar to ours as it combines concrete program execution with a 
symbolic analysis. However, DART applies only to sequential programs, not to 
concurrent programs as we do here. Moreover, DART attempts to cover all the feasible paths 
through the program, not the reachable (abstract) states as we do in our approach. 
DART does not perform any state matching, and therefore it can not detect if an (abstract) 
state has been visited before.
As a result, DART can potentially explore redundant states, e.g. for looping, reactive, programs. 
Another (methodological) difference is that DART uses symbolic 
evaluation while our method uses predicate 
abstraction with refinement.

Dataflow and type-based analyzes have been used to check safety properties of
software (e.g.~\cite{CGS}). Unlike our work, these techniques analyze
over-approximations of system behavior and may generate false reports
due to infeasible paths.

\section{Conclusions and Future Work}
\label{sec:conclusions}

We presented a model checking algorithm based on refinement of
under-approximations, which effectively preserves the defect detection ability
of model checking in the presence of powerful abstractions. The
under-approximation is obtained by traversing the concrete transition system
and performing the state matching on abstract states computed by predicate
abstraction. The refinement is done by checking exactness of abstractions with
the use of a theorem prover. We illustrated the application of the
algorithm for checking safety properties of concurrent programs.
In the future, we plan to investigate whether we can extend the algorithm 
with property driven refinement and with checking liveness properties. 
We also plan to investigate the integration of our approach with 
over-approximation based abstraction refinement and to do an extensive 
evaluation on large systems. 

\section*{Acknowledgement}
We thank the anonymous reviewers for their detailed comments that helped 
us to improve this article significantly.

\end{document}